\documentclass[conference]{IEEEtran}
\ifCLASSINFOpdf
\else
\fi
\usepackage{xcolor}

\begin{document}

\definecolor{FlosColor}{rgb}{0.1,0.8,0.4}
\newcommand{\florian}[1]{\textsf{\textbf{\textcolor{FlosColor}{[TODO: #1]}}}}
%
\title{Self-Reflection as a Tool to Foster Profound Sustainable Consumption Decisions}

\author{\IEEEauthorblockN{Florian Bemmann}
\IEEEauthorblockA{LMU Munich\\
Munich, Germany\\
florian.bemmann@ifi.lmu.de}
\and
\IEEEauthorblockN{Heinrich Hussmann}
\IEEEauthorblockA{LMU Munich\\
Munich, Germany\\
hussmann@ifi.lmu.de}
}


%


\maketitle

\begin{abstract}
 The production of goods we buy on a daily basis accounts for a large portion of greenhouse gas emissions. Although consumers have the power to influence industries' behavior through their demand, making sustainable purchases is challenging.
Current ICT systems supporting sustainable shopping decisions are not established in consumers’ daily lifes. Shopping decisions are made on a complex set of criteria, thus classical persuasive approaches, like recommender-systems, might not be suitable. This work compiles the state of research on ICT supporting sustainable consumption, outlines unsolved challenges, and finally presents a novel concept: a system based on self-reflection instead of classical persuasive approaches, like recommender-systems. Self-reflection provokes revising individual behaviour and decisions, instead of presenting instructions. Combined with additional information on e.g., decision impact, people could learn how to make more sustainable decisions independently.
We envision the deployment of such a system, fostering a change towards more sustainable industries to combat climate change.
\end{abstract}


%
\IEEEpeerreviewmaketitle

\section{Introduction}
The fabrication of goods we consume every day accounts for a large share of our personal, environmental impact \cite{tukker2006environmental,berners2012relative}. Although recently there is a trend towards sustainable production chains, manufacturers strive to keep their production costs low in order to compete. Fortunately, consumers have the power to lead companies to sustainable products and processes, by shifting demand on the market.
This is easier said than done: Among people with an attitude towards sustainable consumption, their behavior often is not \cite{vermeir2004sustainable, hughner2007organic, young2010sustainable}. The barriers to sustainable consumption are mainly informational, despite higher prices being the most relevant barrier:  lack of information about the impact of a product/company, skepticism of certifications, unclear understanding of certifications, a lack of (perceived) availability and insufficient marketing \cite{young2010sustainable,hughner2007organic,aschemann2014elaborating}.

There exist various approaches aiming to overcome the attitude-behavior gap in the domain of sustainable consumption: 
In-shop decision support systems provide information on specific products, e.g. food miles \cite{kalnikaite2011nudge} or consumer-generated environmental impact information \cite{tomlinson2008prototyping, montiel2017mobile}. Recommender-systems were developed to recommend sustainable products \cite{tomkins2018recommender}. However, shopping decisions are subject to complex constraints like family dynamics and daily routines \cite{clear2015supporting,clear2016bearing}, e.g. the distance to the shop, preferences of all family members and required cooking effort. Related systems have to fit into the broader context of life in order to be used. Thus, recommender-systems that decide based on a limited set of criteria are no ideal solution \cite{kalnikaite2011nudge}.
Instead, people should be enabled to decide in a sustainable manner on their own.

Designing for the provision of information on specific products accompanies the problem that such concrete information are not always available \cite{montiel2017mobile}. Furthermore, systems tailored to one specific shop are limited in their applicability, and with recommender-systems users face the difficulty of integrating their decision into everyday life \cite{kalnikaite2011nudge}. 


In this work we propose a system supporting sustainable consumption, based on solutions proposed in existing research and identified drawbacks: Instead of recommending concrete products with quantitative impact data, people should be enabled to make profound decisions themselves. We propose a retrospective analysis of shopping behavior, by self-reflecting on the bought goods in the context of sustainability aspects. In an interactive manner, people should become informed about sustainability aspects regarding their bought product types, and thereby learn to incorporate these insights into their future shopping decisions. Related approaches have been presented by Katzeff et al. who evaluated their dashboard on organic food purchases \cite{katzeff2019encouraging}, a study of self-reflection in the domain of food waste by Ganglbauer et al. \cite{ganglbauer2015and}, or in the Ecofriends app where users reflect on seasonality of foods \cite{tholander2012but}. 


\section{Application Concept}
In this section we present a concept for a novel self-reflection tool that supports sustainable consumption. It arose from existing research in the fields of decision support and information presentation for sustainable behavior, and self-reflection. Our concept is based on the assumptions that (1) people are partially unknowing about how to buy sustainably, (2) manufacturers will produce more sustainably if consumers buy more sustainable products and (3) information on the impact of general product types is available. 

In the following, we describe the key concepts of our system.

\subsection{Enable Profound Decisions instead of Recommending Actions}
To fit into daily life and its constraints, we argue against recommender systems in the domain of consumption decisions.
Instead, people should be enabled to decide in a sustainable manner on their own. Shopping is a repetitive endeavour, that should be influenced by reflection over longer periods of time \cite{clear2016bearing}. The system presented will foster critical reflection of the recent purchase, giving support in the judgement of a product's environmental impact. In the long run people could be enabled to judge products by themselves and might decide more sustainably.

\subsection{Reflection-On-Action}
To allow for a frequent usage, the system should be independent of the supermarket \cite{clear2016bearing}. However, the usage of one's own smartphone is no ideal solution, as it can be overwhelming and lead to poor decisions in the hurry of the supermarket \cite{kalnikaite2011nudge}. Thus we argue for reflection on-action \cite{ploderer2014}: the self-reflection does not take place at the point of time when the decision is made, but afterwards in a more relaxed and flexible environment e.g. on a personal- or tablet computer.

\subsection{Leverage Social- and Group Dynamics}
Families are the main source of constraints for shopping decisions. Hence they should be involved in a sustainability support system \cite{clear2015supporting}. Family members could additionally contribute to the learning process: Relying on an objective information basis about (un-)sustainable characteristics of general product types, interpretation and derivation of consequences concerning the family's cart could be topic of a social process. 

\subsection{Gamification to Trigger Learning about Product Impact}
Gamification can be used to engage people \cite{looyestyn2017does}. In our system it will engage people to learn about the sustainability aspects of their bought products. Thereby, they will understand over time why some product is superior according to a certain criteria set to some other, without prescribing a definition of what is good or bad.

\subsection{Semi-Automatic Purchase Logging}
Manual diary keeping is an important source of self-reflection \cite{ganglbauer2015and}. To lower the burden for users we argue for a semi-automatic diary keeping: the bought products will be captured from the receipt, by taking a photo or through electronic receipts \cite{froehlich2009sensing}. The self-reflection then takes place during data enrichment: users will be engaged to add additional information (e.g. origin country or sustainable impact factor) and be able to correct errors.

\section{How this Work will be presented at the Conference}
Depending on the virtual presentation format, we will present a digital prototype mockup alongside the poster. Visitors can empathize in a typical user story and try the prototype. Screens of the future system will be sketched so that participants can navigate through the system for the given user story. We thereby hope to gain valuable feedback about the idea in general, its concepts, and the prototypical implementation.




%
\bibliographystyle{IEEEtran}
\bibliography{sample-base}



\end{document}